\documentclass[aps,prl,preprint,superscriptaddress]{revtex4}

\usepackage{epsfig} 

\begin{document}

\title{Defective Vortex Lattices in Layered Superconductors with 
Both Point and Correlated  Pins}
\author{J.~P.~Rodriguez}
\affiliation{Dept. of Physics and Astronomy, California State University,
Los Angeles, California 90032, USA.}
%, USA} 

\date{\today}

\begin{abstract}
The mixed phase of layered superconductors with no magnetic screening
is studied through a partial duality analysis of the corresponding
$XY$ model in the presence of random pinning centers.
Sufficently weak Josephson coupling between adjacent layers
results in an entangled vortex solid that exhibits weak superconductivity
across layers.  The corresponding vortex liquid
state shows an inverted specific heat anomaly that is a precursor
to the weak superconductor.\vfill
\bigskip\bigskip\bigskip
keywords :  vortex-lattice pinning, dimensional cross-over, fishtail.
\end{abstract}

\maketitle

\section{Introduction}
\label{intro}
The possibility that the vortex
lattice of a pristine layered superconductor with external magnetic 
field oriented
perpendicular to the layers decouples into an incoherent stack of
two-dimensional (2D) vortex lattices before it melts
has been proposed many times in the 
literature\cite{supersolid}\cite{ffh}\cite{jpr00}.
Monte Carlo simulations and a duality analysis of the corresponding
layered $XY$ model with uniform frustration that describes the
extreme type-II limit find that the decoupled vortex lattice state does
not exist in practice\cite{koshelev97}\cite{jpr02},
however.  The presence of random pinning centers\cite{larkin}\cite{bulaevskii} 
result in quenched-in entanglements of the vortex lines
that can reverse this conclusion\cite{jpr03}.

In this paper, we shall show that random point pins drive a dimensional
cross-over transition in the zero-temperature limit between defective vortex
lattice states that exhibit weak versus strong superconductivity across 
layers.  The  magnetic field at the dimensional cross-over 
transition notably decreases with an increase
in the level of random point pinning. 
This is consistent with recent experimental determinations of
how the ``fishtail'' feature that is  shown by the irreversible magnetization 
of organic superconductors in magnetic field oriented perpendicular 
to the layers
moves with a change in the level of  material disorder\cite{organic}.
We also find that the introduction of correlated pins can act to
increase the dimensional  cross-over field in the zero-temperature limit.
A similar effect has been recently shown  by   the dimensional
cross-over field extracted from flux creep dynamics in high-temperature
superconductors\cite{creep}.
Last, we find that the  zero-temperature cross-over in magnetic field
is inextricably linked  
to the existence of  an inverted specific heat anomaly
within the vortex liquid phase\cite{jpr03}.  The latter accounts for
the experimental observation of a similar thermal anomaly inside the
vortex liquid phase of the high-temperature superconductor YBCO\cite{bouquet}.

\section{Theory}
\label{teoria}

We shall describe the mixed phase of layered superconductors with
random pins by the corresponding $XY$ model with uniform frustration
over the cubic lattice.
Both fluctuations of the magnetic induction and
of the magnitude of the superconducting order parameter are neglected
within this approximation.
The model is therefore valid deep inside
the interior of the mixed phase.  The thermodynamics of the
%three dimensional
three-dimensional (3D) $XY$ model
with anisotropy and with uniform frustration
is determined by the superfluid kinetic energy
\begin{equation}
 E_{XY}^{(3)} = -\sum_{\mu = x, y, z} \sum_r  J_{\mu} {\rm cos}
[\Delta_{\mu}\phi  - A_{\mu}]|_r ,
\label{3dxy}
\end{equation}
which is a  functional of the superconducting
phase $\phi(r)$ over the cubic lattice.
The local phase rigidities within layers, $J_{x}$ and $J_{y}$,
are assumed to be constant over most of the
nearest-neighbor links,
with the exception of those links in the vicinity of a pinning site,
while the Josephson coupling $J_z$ on links between adjacent
layers is assumed to be constant and weak.
The   vector potential
$A_{\mu} = (0, 2\pi f x/a, -b_{\parallel} x)$
represents the magnetic induction,
with a component 
$B_{\perp} = \Phi_0 f / a^2$
oriented perpendicular to the layers,
and with a component
$B_{\parallel} =  (\Phi_0/2\pi d) b_{\parallel}$
oriented parallel to the layers.
Here $a$ denotes the square lattice constant for each layer,
which is of order
the zero-temperature coherence length,
while $d$ represents the separation between layers.
Also,  $\Phi_0$ denotes the flux quantum, 
and $f$ denotes the concentration of planar vortices per site.

{\it Partial Duality Analysis.}
The layered $XY$ model (\ref{3dxy}) can be effectively analyzed in
the selective high-temperature limit, $k_B T\gg J_z$,
through  a {\it partial} duality transformation\cite{jpr00}.
This leads to a dilute Coulomb gas (CG) ensemble
 that describes the nature of the Josephson coupling in terms
of  dual charges that live on the vertical links.
In particular,
thermodynamic quantities are obtained by taking
appropriate derivatives of the
logarithm of the partition function
\begin{equation}
Z_{XY}^{(3)}[0] =  [I_0 (J_z/k_B T)]^{{\cal N}^{\prime}}
\cdot  Z_{\rm CG}[0]
\cdot \Pi_{l} Z_{XY}^{(2)}[0]
\label{z_xy}
\end{equation}
for the layered superconductor,
expressed here as a product of the partition function
for a layered CG ensemble\cite{jpr00}
\begin{equation}
  Z_{\rm CG}[0] = \sum_{\{n_{z}(r)\}} y_0^{N[n_z]}\cdot
\Pi_{l} C_l [q_l]
\cdot e^{-i\sum_r n_z A_z},
\label{z_cg}
\end{equation}
with the contributions from each layer $l$ in isolation, $Z_{XY}^{(2)}$,
along with a modified Bessel function, $I_0 (x)$, raised to the total
number of links between adjacent layers, ${\cal N}^{\prime}$.
Above, the dual charge, $n_z (\vec r, l)$, is an integer field
that lives on links between adjacent layers $l$ and $l+1$
located  at 2D points $\vec r$.
The CG ensemble is weighted
by a product
of phase auto-correlation functions
\begin{equation}
C _l  [q_l] =
\Bigl\langle {\rm exp} \Bigl[i \sum_{\vec r}
q_l (\vec r) \phi (\vec r, l)\Bigr]\Bigr\rangle_{J_z = 0}
\label{c_l}
\end{equation}
for isolated layers $l$
probed at the dual   charge  that accumulates onto
that layer:
\begin{equation}
 q_l (\vec r) = n_z (\vec r, l-1) - n_z (\vec r, l).
\label{q_l}
\end{equation}
It is also weighted
by a bare fugacity
$y_0$   that is
raised to the power
$N [n_z]$
equal to the total
 number of dual charges, $n_z = \pm 1$.
The fugacity is
given by
%$y_0 = J / 2 \gamma^{\prime 2} k_B T$ in the weak-coupling limit
$y_0 = J_z / 2 k_B T$ in the selective high-temperature regime,
$J_z \ll k_B T$, 
reached at large model  anisotropy.
% parameters, $\gamma^{\prime}\rightarrow\infty$,

%{\it Disorder-driven Dimensional Crossover.}
{\it Defective Vortex Lattices.}
A recent analysis of the 2D limit
of the $XY$ model (\ref{3dxy}), $J_z = 0$,
finds that a phase coherent vortex lattice
state with quenched-in unbound dislocations\cite{chudnovsky}
exists in the zero-temperature
limit if the random pinning is weak enough\cite{jpr03}.
This defective vortex lattice state exhibits
2D phase auto-correlation functions (\ref{c_l}) of the form 
$C_l [q] = |C_{l}[q]| {\rm exp} [i\sum_{\vec r} q(\vec r)  \phi_0(\vec r, l)]$,
where the magnitude decays algebraicly  like
\begin{equation}
|C_l [q]|
    = g_0^{n_+}\cdot {\rm exp}\Bigl[  \eta_{2D}
\sum_{(1,2)} q(1){\rm ln} (r_{12} / r_0)\, q(2)\Bigr]
\label{clq}
\end{equation}
asymptotically, 
and where
$\phi_0 (\vec r, l)$ represents the zero-temperature configuration
of layer $l$ in isolation.
% limit, $r_{12}\rightarrow\infty$,
Above, $\eta_{2D}$ is the correlation exponent
that vanishes linearly with temperature in the 
zero-temperature limit,
and   $r_0$ is the natural ultraviolet scale.
% = \eta_{\rm sw} + \eta_{\rm vx}$
%that has  spin-wave and    vortex components given
% by
%$\eta_{\rm sw} = k_B T / 2\pi J_0$  and
%%
%\begin{equation}
%\eta_{\rm vx} = \pi \Bigl\langle
%\Bigl[\sum_{\vec R}^{\qquad\prime} \delta\vec u \Bigr]^2\Bigr\rangle /
%N_{\rm vx} a_{\rm vx}^2,
%\label{etavx1}
%\end{equation}
%%
%respectively.
%The latter monitors fluctuations of the
%center of mass of the vortex lattice,\cite{jpr01} where
%$\delta\vec u$
%% = \vec u - \vec u_0$
% is the diplacement field of
%each vortex with respect to its   location at zero temperature.
%Above, $N_{\rm vx}$ denotes the number of vortices,
%while $a_{\rm vx} = a / f^{1/2}$.
%%is equal to the square root of the area per vortex.
Also , $g_0$ is equal to the ratio of the 2D  phase rigidity
with its value at zero temperature,
while  $n_+$ is equal
to half the number of probes in $q (\vec r)$.
At this stage it becomes useful to re-express the layered CG enesemble
(\ref{z_cg}) by replacing
$C_l [q]$       with its magnitude (\ref{clq}), and by compensating
this change with the additional replacement of
$-A_z (\vec r, l)$ with
\begin{equation}
 \phi_{l, l+1}^{(0)} (\vec r) =
    \phi_0 (\vec r, l+1) - \phi_0(\vec r, l) - A_z (\vec r, l).
\label{phi_0}
\end{equation}
A Hubbard-Stratonovich transformation
of the CG partition function (\ref{z_cg})
then reveals\cite{jpr97} that it is
equal to the partition function
for a   renormalized Lawrence-Doniach (LD) model,
$Z_{\rm LD}[0] = \int {\cal D} \theta\,  e^{-E_{\rm LD}/k_B T}$,
up to a factor that is independent of the Josephson coupling, $J_z$.
The LD energy functional is given by\cite{jpr00}
\begin{equation}
E_{\rm LD} =
\rho_s^{(2D)}\int d^2 r
\sum_{l} \Biggl[
{1\over 2}(\vec\nabla\theta_l)^2
-\Lambda_0^{-2}
{\rm cos}\, \theta_{l, l+1} \Biggr],
\label{e_ld}
\end{equation}
where
$\theta_{l, l+1}  = \phi_{l, l+1}^{(0)} +  \theta_{l+1} - \theta_l$.
The 2D phase rigididity above  is related to the 2D correlation exponent by
$\rho_s^{(2D)} = k_B T / 2\pi\eta_{2D}$.		
Also, $\Lambda_0 = \gamma^{\prime} a$ is the Josephson
penetration depth,
where $\gamma^{\prime} = (J/J_z)^{1/2}$ 
is the model anisotropy parameter with respect to the macroscopic 
phase rigidity of an isolated layer at zero-temperature, $J$.
A standard thermodynamic analysis of the product of partition functions
(\ref{z_xy}) then yields that
the strength of the local Josephson coupling is
given by
$\overline{\langle {\rm cos}\, \phi_{l,l+1}\rangle}  = y_0
 +  g_0 \overline{\langle {\rm cos}\, \theta_{l, l+1} \rangle}$.
Here 
\begin{equation}
 \phi_{l, l+1} (\vec r) =
\phi (\vec r, l + 1) - \phi (\vec r, l) - A_z (\vec r, l)
\label{deltaphi}
\end{equation}
is the gauge-invariant phase difference across adjacent layers,
while the overbar notation represents a bulk average.
It can also be shown\cite{jpr00}
that the  phase rigidity across a macroscopic number of layers is equal to
$\rho_s^{\perp} / J_{z} \cong
g_0 \overline{\langle {\rm cos}\, \theta_{l, l+1} \rangle}$.

To compute  the bulk average
$\overline{\langle {\rm cos}\, \theta_{l, l+1} \rangle}$
at low  temperature,
we must first determine
the configuration $\theta_l^{(0)} (\vec r)$
that optimizes $E_{\rm LD}$.
Eq. (\ref{e_ld}) implies that it satisfies the
field equation
\begin{equation}
-\nabla^2 \theta_l^{(0)} +\Lambda_0^{-2}[{\rm sin}\, \theta_{l-1, l}^{(0)}
- {\rm sin}\, \theta_{l, l+1}^{(0)}] = 0.
\label{fldeqs}
\end{equation}
In the weak-coupling limit,  $\Lambda_0\rightarrow\infty$,
we therefore have that
$\theta_l^{(0)} (\vec r)$  is constant inside of each layer.
Also, by analogy with 2D melting physics\cite{jpr01}\cite{NH},
the presence of quenched-in unbound dislocations
in the pinned vortex lattice\cite{jpr03}\cite{chudnovsky}
implies that
the quenched-in  inter-layer order parameter 
${\rm exp}\, i\phi_{l, l+1}^{(0)}$
shows only short-range correlations on average over the bulk.
Specifically, we have
\begin{equation}
\overline{{\rm exp} [i\phi_{l, l+1}^{(0)} (1)]
\cdot
{\rm exp} [-i\phi_{l, l+1}^{(0)} (2)]}
= e^{i b_{\parallel} x_{12}} \, e^{-r_{12}/l_{\phi}}
\label{defl_phi}
\end{equation}
asymptotically,
where  $l_{\phi}$ is a 2D 
%zero-temperature
disorder  scale set by the density of unbound
dislocations between adjacent layers (see next section).
%The absence of long-range correlation in the inter-layer
%order parameter ${\rm exp}\, i\phi_{l, l+1}^{(0)}$ on average
The absence of long-range order 
along adjacent layers on  average over the bulk
then implies that
$\overline{{\rm cos}\, \theta_{l, l+1}^{(0)}}
\cong  \overline{(\delta\theta_{l}^{(0)} - \delta\theta_{l+1}^{(0)})
\cdot {\rm sin}\,\theta_{l, l+1}^{(0)}}$
at zero temperature in the weak-coupling limit\cite{larkin},
where $\delta\theta_l^{(0)}$ is the deviation
with respect to
the constant background.
This yields the perturbative result
\begin{equation}
\overline{{\rm cos}\, \theta_{l, l+1}^{(0)}}
\cong
\Lambda_0^{-2} \int d^2 r_{12}\,
 \overline{{\rm exp}[i\phi_{l, l+1}^{(0)} (1)]
\cdot
{\rm exp} [-i\phi_{l, l+1}^{(0)} (2)]}
\cdot G^{(2)} (r_{12})
\label{pert}
\end{equation}
at weak coupling by Eq. (\ref{fldeqs}),
where $G^{(2)} = - \nabla^{-2}$ is the 2D Greens function.
%At weak coupling,
After taking the separation at which $G^{(2)}$ vanishes to be of
order $\Lambda_0$,
we then obtain the final perturbative result\cite{bulaevskii}
\begin{equation}
\overline{{\rm cos}\, \theta_{l, l+1}^{(0)}}\cong
[A_0 - A_2 (b_{\parallel} l_{\phi})^2]
(l_{\phi}/\Lambda_0)^2 {\rm ln}(\Lambda_0 / l_{\phi})
\label{LDcos}
\end{equation}
for the bulk average 
at  zero temperature in the weak-coupling limit,
and  at weak parallel field $B_{\parallel}\ll\Phi_0/2\pi l_{\phi} d$.
Here $A_0$ and $A_2$ are numerical constants of order unity:
e.g., $A_0 = 1$ and $A_2 = 3/2$ if Eq. (\ref{defl_phi}) is generally valid.
By the discussion following Eq. (\ref{deltaphi}),
we conclude that
random point pins result in a vortex glass
that exhibits weak superconductivity
%a relatively low phase rigidity and ``cosine''
across layers\cite{jpr00},
$\rho_s^{\perp} \ll J_z$,
at sufficiently high layer  anisotropy\cite{jpr03},
%\cite{larkin}\cite{bulaevskii},
$\Lambda_0\gg l_{\phi}$.
%and $\overline{\langle{\rm cos}\, \phi_{l, l+1}\rangle}\ll 1$.
Notice, however, that the quadratic dependence with small parallel field
shown by the interlayer LD cosine (\ref{LDcos})
implies the {\it absence} of line tension for Josephson vortices!
We believe that this reflects the vortex-glass nature of the
phase coherence\cite{ffh}:
$\overline{\langle e^{i\phi_{l,l+1}(0)} e^{-i\phi_{l,l+1}(\infty)}\rangle} = 0$
and 
$\overline{|\langle e^{i\phi_{l,l+1}(0)} e^{-i\phi_{l,l+1}(\infty)}\rangle|^2}
> 0$.
Last, the bulk average at zero temperature
$\overline{{\rm cos}\, \theta_{l, l+1}^{(0)}}$
must pass between zero and unity 
at   $\Lambda_0\sim l_{\phi}$ 
as a function of the anisotropy parameter.
This condition then defines a decoupling cross-over field
\begin{equation}
B_D^{\perp} (0) \sim ( l_{\phi}/a_{\rm vx})^2 (\Phi_0/\Lambda_0^2),
\label{xover}
\end{equation}
at which point the reversible magnetization shows a broad
diamagnetic peak.  Above, $a_{\rm vx} = a/f^{1/2}$ is
the planar inter-vortex scale.

%{\it Inverted Specific-heat Anomaly.}
{\it Decoupled Vortex Liquid.}
Suppose now that the macroscopic
phase coherence (\ref{clq}) shown by each layer in isolation
at low  temperature
is lost at temperatures above  a transition temperature $T_g^{(2D)}$,
and that only short-range phase correlations
over a scale $\xi_{2D}$ exist.
A high-temperature expansion\cite{koshelev96}
of the CG ensemble (\ref{z_cg}) in powers of the fugacity $y_0$ 
is then possible at temperatures above the 2D glass transition 
in the weak-coupling limit.
At zero parallel field in particular,  Eqs. (\ref{z_xy}) and (\ref{z_cg})
yield the expansion 
\begin{eqnarray}
\overline{\langle {\rm cos}\, \phi_{l, l+1}\rangle} \cong  
   &y_0&\sum_1 \overline{C_l (0,1) \cdot C_{l+1}^* (0,1)} - \nonumber \\
 &y_0^3&\Biggl[\sum_{1,2,3} \overline{C_l (0,1) \cdot C_{l+1}^* (0,1)
\cdot
C_l (2,3) \cdot C_{l+1}^* (2,3)}
-
\sum_{1,2,3}^{\qquad\prime} 
\overline{C_l (0,1,2,3) \cdot C_{l+1}^* (0,1,2,3)}\Biggr]  \nonumber \\
\label{cos1}
\end{eqnarray}
for the inter-layer ``cosine'' to next-leading order in the fugacity.
The prime over the sum in the last term above excludes double counting
of probes $(0, 1, 2, 3)$.
Also, next-leading-order contributions that couple together 
{\it three} adjacent layers are negligible in the critical regime,
$\xi_{2D}\gg l_{\phi}$,
and these are not listed above.
Substitution of the  form
$C_l (1,2) = g_0 e^{-r_{12}/\xi_{2D}} e^{i\phi_0 (1)} e^{-i\phi_0 (2)}$
for the   phase autocorrelation function (\ref{c_l})
into the leading term above 
%The overbar represents a disorder (sample) average,
yields a  non-divergent result
\begin{equation}
\overline{\langle {\rm cos}\, \phi_{l, l+1}\rangle} \sim
g_0^2 (J / k_B T)   [(l_{\phi}^{-1} + \xi_{\phi}^{-1})^{-1} / \Lambda_0]^2,
\label{cos2}
\end{equation}
where $\xi_{\phi} = \xi_{2D} / 2$.
%
%that is valid in the decoupled vortex liquid that exists at
%fields much larger than $f\gamma_{\times}^{\prime 2}$.
%Last, Eq. (\ref{cos2})
This 
implies an anomalous inter-layer contribution to the specific heat per volume
%vertical link
%$c_{v}^{\perp} = -\partial e_J / \partial T$,
 equal to
\begin{equation}
\delta c_{v}^{\perp}
%^{\perp}
\cong
%  e_J^{(0)} \overline{\langle {\rm cos}\, \phi_{l, l+1}\rangle}
 2 [ 1 + (\xi_{\phi} /  l_{\phi})]^{-1}
(\partial {\rm ln}\, \xi_{\phi}^{-1} /  \partial T ) e_{J},
\label{c_perp}
\end{equation}
where
$e_{J} = \overline{\langle {\rm cos}\, \phi_{l, l+1}\rangle}
\cdot J/\Lambda_0^2 d$
%$e_J^{(0)} = J/\Lambda_0^2 d$
%$e_J^{(0)} = J_z/a^2 d$
is the
%bare
Josephson energy density,
and where  $d$ denotes the spacing in between adjacent layers.
It notably shows an inverted
specific heat jump that is followed by a  tailoff
at a temperature $T_p$ such that
$\xi_{\phi} (T_p) \sim  l_{\phi}$
% the 2D  phase correlation length
% $\xi_{2D} (T_p)$
%matches the disorder scale $l_{2D}$
if $\xi_{\phi}$ diverges faster
than $[T - T_g^{(2D)}]^{-1}$.

The next-leading-order term for the
inter-layer ``cosine''
(\ref{cos1}) is negative,
divergent, and of order
$-y_0^3 g_0^4 l_{\phi}^4 \xi_{\phi}^2 / a^6$
in the critical region, $\xi_{\phi} \gg l_{\phi}$.
This  means that it
becomes comparable to the leading
order term (\ref{cos2})
at magnetic fields below a   
%2D-3D 
cross-over scale
\begin{equation}
B_{\times}^{\perp}
\sim g_0 (J/k_B T) (l_{\phi} \xi_{\phi}/a_{\rm vx}^2) (\Phi_0/\Lambda_0^2)
\label{2d-3d}
\end{equation}
that separates 2D from 3D vortex-liquid behavior\cite{jpr02}\cite{jpr03}.
The phase correlation length across layers grows larger than
the separation between adjacent layers at temperatures below
this point, $T_{\times}$. 
It then diverges (or jumps to infinity) at a 
3D ordering temperature, $T_g$, that lies inside of the window
$[T_g^{(2D)}, T_{\times}]$.
The phase diagram displayed by Fig. 1 summarizes all of the above.

\section{Discussion and Conclusions}
\label {disc}
The present theory predicts a disorder-driven dimensional crossover 
(\ref{xover})
between defective vortex lattices that show weak versus strong
superconductivity across layers.
Comparison of the bulk average (\ref{defl_phi})  
of the quenched-in phase autocorrelation
with the corresponding thermal average in
the hexatic phase of a pristine vortex liquid\cite{jpr01}
indicates that the 2D disorder scale, $l_{\phi}$, that controls the 
crossover transition
is set by the density of unbound
dislocations {\it between} adjacent layers
in the vortex lattice.  For example,
the addition or the removal of a straight line of dislocations
that traverses all of the
layers will not change   this scale.	A useful 
parameterization of $l_{\phi}$ is then
\begin{equation}
l_{\phi} = l_{2D}/(2\, x_{\rm point}^{1/2}),
\label{l_phi}
\end{equation}
where $l_{2D}$ is the correlation length of the quenched-in order parameter
${\rm exp}(i\phi_0)$ on average over a  given  layer in isolation, 
and where $x_{\rm point}$ is the concentration of unbound dislocations
{\it between} adjacent layers relative to the total number of unbound
dislocations in adjacent layers: e.g., $x_{\rm point} = 0$ if all
of the unbound dislocations form smooth lines across layers,
while $x_{\rm point} = 1$ if sections of such  lines are absent.
Eqs. (\ref{xover}) and (\ref{l_phi}) then yield two important predictions:
the decoupling crossover field $B_D^{\perp} (0)$ is 
({\it i}) diminished  
by    an increase in the 
density of unbound dislocations between adjacent layers, $l_{\phi}^{-2}$,
and it ({\it ii}) increases with an increase in the concentration of
unbound dislocations that line-up across
adjacent  layers, $1 - x_{\rm point}$.
Prediction ({\it i}) is consistent with a recent proposal that  the
second magnetization peak
(or ``fishtail'')  observed in the mixed phase of organic
superconductors is due to  dimensional  crossover
of the vortex lattice\cite{organic}.
The second-peak shown by the irreversible magnetization 
that is measured in  this work 
%ref. \cite{organic} 
moves to lower perpendicular magnetic
field as the  concentration  of random point pins increases.  
Also, prediction ({\it ii}) is  consistent 
with recent experiments that observe
a 2D-3D crossover in the flux creep dynamics 
of high-temperature superconductors in 
magnetic field oriented perpendicular  to the layers\cite{creep}.
%In particular
The dimensional cross-over field that is extracted experimentally
in ref. \cite{creep} is
found to increase after columnar pins are introduced.  It
is quite possible that these help to line-up a fraction of  the existing
unbound dislocations across layers, which in turn leads  
to an increase in the decoupling field
(\ref{xover}) as a result of the increase in the 
2D disorder scale $l_{\phi}$.
% by Eq. (\ref{l_phi}).

The second important result obtained here is the inverted
specific-heat anomaly, Eq. (\ref{c_perp}),
that is predicted to exist inside of the decoupled
vortex liquid phase.  Recent experiments
on 2D arrays of Josephson junctions with bond disorder
in external magnetic field are
consistent with a continuous vortex glass transition
that has a 
2D phase correlation length that diverges
as $\xi_{2D} \propto [T - T_g^{(2D)}]^{-\nu_{2D}}$
at the transition,
with a critical exponent $\nu_{2D} = 2$ \cite{arrays}.
This implies an inverted 
specific heat anomaly, Eq. (\ref{c_perp}),
that peaks  at $\xi_{2D} (T_p) = 2 l_{\phi}$, 
which  is shown in Fig. 2.
A similar anomaly was observed within the vortex-liquid phase
of YBCO\cite{bouquet}.  Comparison of the magnitude of that anomaly
with Eq. (\ref{c_perp}) yields a 10\% change in the inter-layer
cosine (\ref{cos2}) across the 
vortex-liquid/vortex-liquid transition\cite{jpr03}.
The above indicates that the specific-heat anomaly observed
experimentally inside of  the vortex-liquid phase of YBCO
represents not a true phase transition\cite{tesanovic},
but a crossover instead.

Noticably absent from the above calculations is the effect of
magnetic coupling between planar vortices across layers.
This in general reduces the entanglement of vortex lines across layers,
and it will therefore increase the  inter-layer phase coherence,
Eqs. (\ref{LDcos}) and (\ref{cos2}).  Ref. \cite{bulaevskii}
shows, however, that the  effect of 
magnetic coupling can be accounted for by
a renormalization up in the 2D disorder scale, $l_{\phi}$.
Last, a sharp and possibly first-order disorder-driven  decoupling
transition has been found numerically at zero temperature
for the vortex lattice with magnetic interactions, but
with no Josephson coupling\cite{olson}.
This indicates that the disorder-driven dimensional crossover found here
in the zero-temperature limit survives the addition of 
weak magnetic coupling.

We conclude that random point pinning of the vortex
lattice in layered superconductors results in a 2D-3D crossover
transition at low  temperature that separates
strong from  weak superconductivity across layers (see Fig. 1).
The disorder-driven phenomenon suggests the following
physical picture: although the tension for a line of
dislocations in the defective vortex lattice is null on both sides of
the crossover, the scale of the roughness in the line
is large compared to the inter-layer spacing in the 3D 
phase, while it  is of order
or less than the inter-layer spacing in the quasi-2D phase.
Scaling suggests that such quenched-in disorder is established
first at separations
between layers of order $l_{\phi}/\gamma$, where 
$\gamma = \Lambda_0/d$ is the physical anisotropy parameter.

The author thanks C. Reichhardt and C. Olson for informative discussions.

%\begin{turnpage}
\begin{figure}
\includegraphics[scale=0.36, angle=-90]{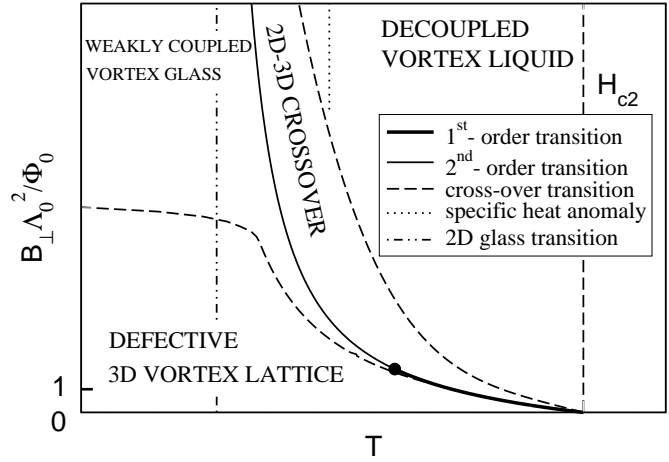}
%\centerline{\epsfxsize=60mm \epsfbox{FG31.ps}}
\caption{Shown is a proposed phase diagram assuming weak point pins
and a continuous vortex glass phase transition for isolated layers.
The perpendicular magnetic field is held fixed while the Josephson
penetration depth, $\Lambda_0$, is swept along the verticle axis.
A   mean-field temperature dependence,
$J\propto T_{c0} - T$, is also assumed.}
\label{phasedia}
\end{figure}  
%\end{turnpage}

\begin{figure}
\includegraphics[scale=0.36, angle=-90]{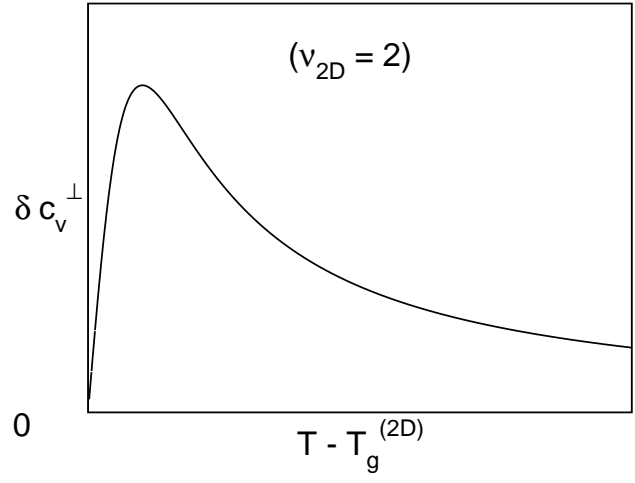}
\caption{Theoretical specific heat anomaly, Eq. (\ref{c_perp}),
for $\xi_{2D}\propto [T - T_g^{(2D)}]^{-2}$.
The peak occurs at $\xi_{2D} (T_p) = 2 l_{\phi}$.}
\label{anomaly}
\end{figure}

\end{document}